# Post-Quantum Cryptography: Securing Digital Communication in the Quantum Era

Dr. G S Mamatha
*Professor*
R.V. College of Engineering
Bengaluru, India.PIN: 560059
Email: mamathags@rvce.edu.in

Rasha Sinha
*B.E. ISE*
R.V. College of Engineering
Bengaluru, India.PIN: 560059
Email: rashasinha.is21@rvce.edu.in

Namya Dimri
*B.E. ISE*
R.V. College of Engineering
Bengaluru, India.PIN: 560059
Email: namyadimri.is21@rvce.edu.in

*Abstract*—The advent of quantum computing poses a profound threat to traditional cryptographic systems, exposing vulnerabilities that compromise the security of digital communication channels reliant on RSA, ECC, and similar classical encryption methods. Quantum algorithms, notably Shor's algorithm, exploit the inherent computational power of quantum computers to efficiently solve mathematical problems underlying these cryptographic schemes. In response, post-quantum cryptography (PQC) emerged as a critical field aimed at developing resilient cryptographic algorithms impervious to quantum attacks. This paper delineates the vulnerabilities of classical cryptographic systems to quantum attacks, elucidates the principles of quantum computing, and introduces various PQC algorithms such as lattice-based cryptography, code-based cryptography, hash-based cryptography, and multivariate polynomial cryptography. Highlighting the importance of PQC in securing digital communication amidst quantum computing advancements, this research underscores its pivotal role in safeguarding data integrity, confidentiality, and authenticity in the face of emerging quantum threats.

*Keywords—Quantum computing, Shor's algorithm, Cryptographic systems, RSA, Post-quantum cryptography(PCQ)*

## I. Introduction

In an era marked by exponential advancements in computing capabilities, the emergence of quantum computing stands as a paradigm-shifting development with profound implications for cybersecurity. At the heart of this transformation lies the fundamental challenge posed to traditional cryptographic systems, which form the bedrock of secure digital communication. Classical encryption methods, including RSA and ECC, have long been relied upon to safeguard sensitive information exchanged over digital networks. However, the advent of quantum computing brings forth a new frontier of computational power, capable of unraveling the mathematical underpinnings upon which these cryptographic schemes rest. Shor's algorithm, a cornerstone of quantum computing, poses a particularly formidable threat by efficiently factoring large composite numbers and solving discrete logarithm problems, tasks considered intractable for classical computers. Recognizing the vulnerability of traditional cryptographic systems to quantum attacks, the field of post-quantum cryptography emerges as a vital avenue for developing resilient cryptographic algorithms that can withstand the disruptive force of quantum computing. This paper explores the vulnerabilities of classical cryptographic systems to quantum attacks, elucidates the principles of quantum computing, and underscores the importance of post-quantum cryptography in securing digital communication in the face of emerging quantum threats.

## II. Quantum Computing Threats to Cryptography

Quantum computing operates on the principles of quantum mechanics, leveraging the unique properties of quantum bits or qubits to perform computations. Unlike classical bits, which can represent either 0 or 1, qubits can exist in superposition, representing both 0 and 1 simultaneously. Additionally, qubits can be entangled, meaning the state of one qubit is dependent on the state of another, even if they are separated by great distances.

The impact of quantum computing on cryptographic algorithms stems from its ability to execute certain mathematical operations exponentially faster than classical computers. Cryptographic algorithms like RSA and ECC rely on the difficulty of certain mathematical problems, such as integer factorization and discrete logarithms, for their security. However, quantum algorithms, particularly Shor's algorithm, can efficiently solve these problems, posing a significant threat to traditional cryptographic systems

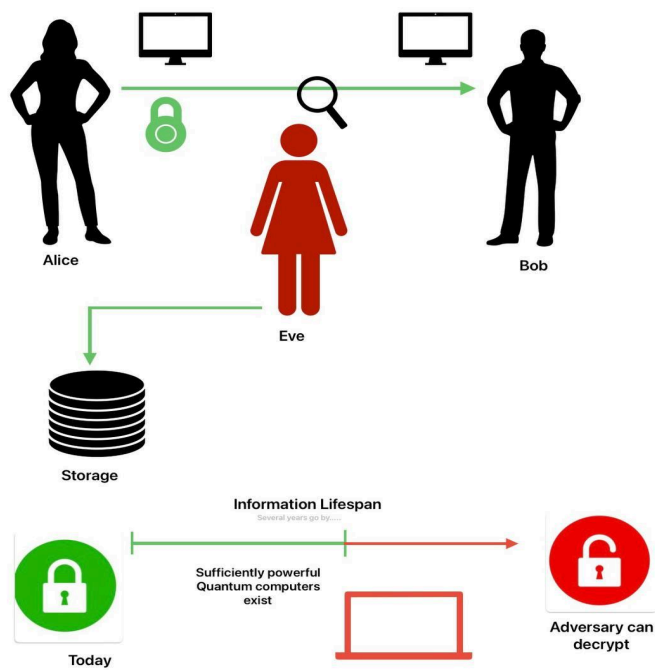

Fig1. Quantum computer attack on RSA

*A. Shor's Algorithm :* Shor's Algorithm stands as a groundbreaking advancement in the realm of quantum computing, particularly due to its capability to efficiently solve two critical mathematical problems: integer factorization and the discrete logarithm problem on elliptic curves. These problems underpin the security of RSA and ECC encryption schemes, making Shor's Algorithm a potent threat to these widely used cryptographic systems.Breaking RSA encryption: RSA encryption, named after its inventors Ron Rivest, Adi Shamir, and Leonard Adleman, relies on the difficulty of factoring large composite numbers into their prime factors. The security of RSA encryption hinges on the practical infeasibility of factoring large integers with classical computers. However, Shor's Algorithm demonstrates the ability to factorize large integers exponentially faster on a quantum computer compared to classical algorithms. This means that Shor's Algorithm can break RSA encryption by efficiently finding the prime factors of the modulus, thus revealing the private key.

*B. Breaking ECC Encryption:* Elliptic Curve Cryptography (ECC) is another widely-used cryptographic scheme that relies on the hardness of the elliptic curve discrete logarithm problem. ECC offers strong security with relatively smaller key sizes compared to RSA. However, the security of ECC is threatened by Shor's Algorithm due to its ability to solve the elliptic curve discrete logarithm problem efficiently on a quantum computer. This means that Shor's Algorithm can compromise the security of ECC by efficiently computing the discrete logarithm, enabling an attacker to derive the private key from the public key.

### III. POST-QUANTUM CRYPTOGRAPHY FUNDAMENTALS

Post-Quantum Cryptography (PQC) encompasses a diverse range of cryptographic algorithms designed to resist attacks from quantum computers. Among the prominent PQC algorithms are lattice-based cryptography, code-based cryptography, hash-based cryptography, multivariate polynomial cryptography, and various others.

- Lattice-based Cryptography: Lattice-based cryptography relies on the hardness of certain problems defined on lattices, such as the Shortest Vector Problem (SVP) or the Learning With Errors (LWE) problem. These problems are believed to be resistant to quantum attacks due to their inherent complexity.

- Code-based Cryptography: Code-based cryptography relies on error-correcting codes, such as Goppa codes or McEliece cryptosystems. The security of these systems is based on the difficulty of decoding certain linear codes, which is conjectured to be resistant to quantum attacks.

- Hash-based Cryptography: Hash-based cryptography utilizes cryptographic hash functions to provide digital signatures and other cryptographic primitives. These schemes are based on the security properties of cryptographic hash functions, which are not directly threatened by quantum attacks.

- Multivariate Polynomial Cryptography: Multivariate polynomial cryptography relies on the hardness of solving systems of multivariate polynomial equations. The security of these schemes is based on the computational complexity of solving systems of equations, which is believed to be resistant to quantum attacks.

*A. Resistance of PQC Algorithms Against Quantum Attacks:* PQC algorithms are designed to be resistant to attacks from quantum computers, which have the potential to break classical cryptographic schemes such as RSA and ECC using algorithms like Shor's algorithm. The resistance of PQC algorithms against quantum attacks stems from their reliance on mathematical problems that are believed to be hard even for quantum computers to solve efficiently. These problems are chosen specifically to resist attacks based on quantum algorithms, ensuring the security of cryptographic systems in the presence of quantum adversaries.

*B. Comparison of PQC Algorithms:* In the wake of quantum computing advancements, the necessity for cryptographic algorithms resilient to quantum attacks has spurred the development of a plethora of Post-Quantum Cryptography (PQC) schemes. This section presents a comparative analysis of prominent PQC algorithms, focusing on security, efficiency, and implementation complexity.

| TYPE | ADVANTAGE | DISADVANTAGES | ALGORITHM |
|---|---|---|---|
| *Lattice* | Fast operation speed | Difficult setting parameter | CRYSTALS-DILITHIUM, FALCON |
| *Code* | Small signature size, Fast operation speed | Large key size | Classic McEliece, BIKE, HQC |
| *Multi-variate* | Fast encryption and decryption speed | Large key size | Rainbow |
| *Isogeny* | Small key size | Slow operation speed | SIKE |
| *Hash* | Safety proof possible | Large signature size | SPHINCS+, PICNIC |

Table 1. Comparison of PQC Algorithms

## IV. Implementation Challenges and Considerations:

The adoption of post-quantum cryptographic (PQC) systems presents several implementation challenges and considerations, reflecting the need for seamless integration into existing cryptographic infrastructures and protocols.

*A. Challenges of Implementing PQC Algorithm:* Implementing post-quantum cryptographic (PQC) algorithms presents several challenges that require careful consideration and strategic planning. One of the primary challenges lies in the algorithmic complexity inherent in many PQC schemes. Unlike classical cryptographic algorithms, PQC algorithms often rely on intricate mathematical structures and novel computational techniques, demanding specialized expertise for successful implementation. Furthermore, the resource intensiveness of certain PQC algorithms poses significant hurdles, as they may require substantial computational resources and memory overhead, potentially straining existing hardware infrastructures. Interoperability issues also emerge as a critical concern, as seamless integration with legacy cryptographic systems and protocols is paramount for ensuring smooth deployment and operation. Additionally, the lack of standardized protocols and certification procedures for PQC algorithms complicates the implementation process, necessitating collaborative efforts among industry stakeholders to establish robust standards and certification frameworks. Addressing these challenges requires comprehensive planning, rigorous testing, and collaboration across multidisciplinary teams to navigate the complexities of implementing PQC algorithms effectively in real-world environments.

*B. Considerations for Transitioning to PQC:*

1) Interoperability Planning: Organizations must develop robust interoperability strategies to facilitate the seamless integration of PQC algorithms with existing cryptographic systems and protocols. Compatibility testing and validation procedures should be conducted rigorously to minimize disruption.

2) Algorithm Selection: Careful consideration must be given to the selection of PQC algorithms, taking into account factors such as security guarantees, performance characteristics, and implementation complexity. Collaborative efforts with cryptographic experts and industry stakeholders can inform prudent algorithmic choices.

3) Legacy System Compatibility: Transitioning from classical to post-quantum cryptographic systems requires compatibility with legacy systems. Implementers should devise comprehensive migration plans, considering phased deployment strategies and backward compatibility measures to mitigate disruption and ensure smooth transition.

4) Compliance and Regulatory Requirements: Organizations must adhere to relevant cryptographic standards and regulatory requirements when implementing PQC algorithms. Close collaboration with regulatory bodies and industry stakeholders is essential to navigate compliance complexities and ensure adherence to legal frameworks.

*C) Scalability and Performance Concerns:* Scalability and performance concerns are paramount in the implementation of post-quantum cryptographic (PQC) algorithms, particularly as organizations seek to deploy these solutions across diverse operational environments. The scalability of PQC algorithms is crucial for accommodating increasing computational demands and accommodating large-scale deployment scenarios. Organizations must evaluate the scalability of PQC solutions to ensure they can effectively handle growing workloads and evolving cryptographic requirements. Additionally, performance optimization becomes a critical consideration, necessitating exploration of algorithmic enhancements, hardware acceleration techniques, and parallel processing strategies to maximize computational efficiency and throughput. Rigorous performance testing and benchmarking are essential to identify potential bottlenecks, optimize system performance, and ensure that PQC algorithms meet stringent performance requirements in real-world applications. By addressing scalability and performance concerns proactively, organizations can enhance the efficiency, reliability, and scalability of their cryptographic infrastructures, thereby strengthening their resilience against emerging quantum threats.

## V. Current Research and Development

Driven by the looming threat of quantum computers, the field of Post-Quantum Cryptography (PQC) has seen significant progress recently. Recognizing the vulnerability of current cryptographic standards to advancements in quantum computing, organizations like the National Institute of Standards and Technology (NIST) have accelerated efforts to identify and implement PQC solutions. This has resulted in the finalization of the first batch of PQC algorithms by NIST, marking a crucial milestone. These algorithms, including CRYSTALS-Kyber for general encryption and CRYSTALS-Dilithium for digital signatures, are specifically designed to resist attacks from quantum computers.

However, the journey doesn't end with selecting algorithms. Standardization efforts are now in full swing, with organizations like NIST and the Internet Engineering Task Force (IETF) actively collaborating to develop protocols and standards for seamlessly integrating PQC algorithms into existing infrastructure. This involves creating guidelines and specifications for secure implementation, ensuring interoperability between different systems, and minimizing disruption during the transition from traditional to PQC-based cryptography. The ultimate goal is to create a future-proof cryptographic landscape that can withstand the challenges posed by quantum computing, while maintaining robust security for sensitive information across various applications.

*A) Standardization Efforts by NIST and ETSI:* NIST(National Institute of Standards and Technology) plays a pivotal role in the standardization of cryptographic algorithms and protocols, including efforts related to post-quantum cryptography (PQC). Their PQC project aims to develop cryptographic standards that can withstand quantum attacks, which threaten current cryptographic systems based on integer factorization and discrete logarithm problems. NIST has solicited proposals for PQC algorithms since 2017 and has been rigorously evaluating submissions for security, performance, and practicality. The selection process involves multiple rounds of analysis, public feedback, and scrutiny to ensure the chosen algorithms meet stringent criteria.

Similar to NIST, ETSI (European Telecommunications Standards Institute) actively participates in standardization efforts within the telecommunications sector, including cryptographic algorithms and protocols. While ETSI does not have a dedicated PQC project like NIST, it collaborates with various stakeholders to ensure that cryptographic standards remain robust and secure against emerging threats, including those posed by quantum computing. ETSI's contributions to standardization efforts often involve coordinating with industry experts, researchers, and policymakers to develop comprehensive standards that address diverse use cases and regulatory requirements within the European Union and beyond.

VI. POST-QUANTUM CRYPTOGRAPHY (PQC) ALGORITHMS AND THEIR POTENTIAL APPLICATIONS:

*A) Lattice-Based Cryptography:* Lattice-based cryptography is a branch of cryptography that relies on the hardness of lattice problems for its security. Lattices are geometric structures formed by points in multidimensional space, and lattice-based cryptography exploits the computational complexity of certain lattice problems to create cryptographic primitives that are resistant to attacks, including those from quantum computers.

In lattice-based cryptography, the security of cryptographic schemes is based on the assumed difficulty of lattice problems, such as the Shortest Vector Problem (SVP) and the Closest Vector Problem (CVP). These problems involve finding the shortest or closest lattice vector to a given point in the lattice and are known to be computationally hard, even for quantum computers.

Lattice-based cryptography, rooted in the computational complexity of lattice problems, provides a versatile and quantum-resistant framework for cryptographic primitives. Algorithms like NTRUEncrypt, Kyber, and NewHope leverage lattice structures to offer secure communication, digital signatures, and key exchange protocols. Its resilience to quantum attacks, along with applications spanning secure communication channels to homomorphic encryption, underscores its significance in post-quantum security efforts and future cryptographic standards. Ongoing research and standardization efforts continue to solidify lattice-based cryptography's role in safeguarding sensitive information and ensuring the integrity of digital transactions amidst the quantum computing era.

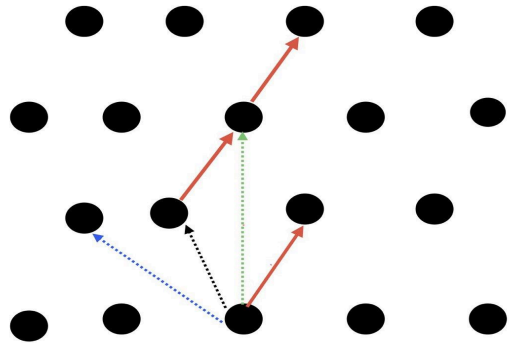

Fig 2. Lithium-based cryptography

1. Versatility: Lattice-based cryptography offers a wide range of cryptographic primitives, including encryption schemes, digital signatures, and key exchange protocols. This versatility allows for the construction of secure cryptographic systems that can be tailored to specific application requirements.

2. Quantum Resistance: Lattice-based cryptography is inherently resistant to attacks from quantum computers. Unlike traditional cryptographic schemes based on number theoretic problems like integer factorization and discrete logarithm, lattice-based schemes do not rely on assumptions that are vulnerable to quantum algorithms like Shor's algorithm.

3. Security Assumptions: The security of lattice-based cryptography relies on the hardness of certain lattice problems. While these problems are believed to be computationally difficult, there is ongoing research to analyze the security of lattice-based schemes and to develop efficient algorithms for solving lattice problems.

*B) Code-based Cryptography:* It is a stalwart in the realm of post-quantum security, relies on the intricate properties of error-correcting codes to fortify cryptographic protocols against quantum threats. Unlike traditional cryptographic methods vulnerable to quantum algorithms, code-based approaches like the McEliece cryptosystem and BIKE algorithm derive their strength from the complexity of error-correcting codes, making them resilient to quantum attacks. While these schemes often entail higher computational costs compared to classical cryptography, their quantum resistance positions them as pivotal solutions for long-term security in an increasingly quantum-capable world. Code-based cryptography finds applications across diverse domains, encompassing secure communication, digital signatures, and key exchange mechanisms, providing a robust shield against emerging quantum threats and ensuring the integrity and confidentiality of sensitive information in the digital age. As research and standardization efforts progress, code-based cryptography is poised to play a foundational role in shaping the future of secure communication and data protection amidst the advent of quantum computing.

*C) Multivariate Polynomial Cryptography (MPC):* It represents a formidable branch of post-quantum cryptography, leveraging the complexity of solving systems of multivariate polynomial equations for cryptographic security. Unlike traditional cryptographic schemes, MPC algorithms like Rainbow and HFE rely on the formidable computational challenges posed by solving systems of multivariate polynomial equations, making them resistant to attacks from both classical and quantum adversaries. These algorithms offer robust security guarantees, even in the face of advances in quantum computing. One of the key strengths of MPC lies in its versatility, allowing it to be applied to various cryptographic primitives such as digital signatures, encryption schemes, and authentication protocols. The inherent mathematical complexity of MPC algorithms forms a solid foundation for security while enabling practical implementation in real-world systems. Moreover, MPC schemes offer an alternative cryptographic approach suitable for specific use cases where other post-quantum cryptography (PQC) algorithms may not be as effective.

In practical terms, Multivariate Polynomial Cryptography finds applications in securing communication channels, digital signatures, and encryption protocols, providing a resilient shield against quantum attacks and ensuring the confidentiality and integrity of sensitive information. As research and development in MPC progresses, it holds the potential to significantly impact the landscape of post-quantum cryptography, offering robust and versatile solutions for securing digital communications and safeguarding critical infrastructure in the era of quantum computing.

*D) Hash Based Algorithm:* Hash-based cryptography within the framework of post-quantum cryptography (PQC) is a crucial area of exploration and development in response to the potential threat posed by quantum computing to traditional cryptographic systems. Unlike classical computing, quantum computers leverage principles of quantum mechanics to perform computations significantly faster than classical computers. This advancement could render many current cryptographic algorithms, such as RSA and ECC, vulnerable to attacks, prompting the need for alternative solutions like hash-based cryptography. In hash-based cryptography, one-way hash functions play a central role. These functions take an input and produce a fixed-size output, known as a hash value or digest. The key property of one-way hash functions is their irreversibility: given a hash value, it is computationally infeasible to determine the original input. This property ensures the security of hash-based cryptographic schemes by making it difficult for adversaries to derive sensitive information from hashed data.

One of the most prominent applications of hash-based cryptography in the context of PQC is in digital signatures. Digital signatures are essential for authenticating the origin and integrity of digital messages or documents. In hash-based signature schemes, the signer generates a hash value of the message and then signs the hash value with their private key. Recipients can verify the signature using the signer's public key and compare it with the hash value of the message they compute independently.

Another notable application of hash-based cryptography is in secure data structures such as Merkle trees. Merkle trees are binary trees constructed using hash values, where each leaf node represents a data block and each non-leaf node represents the hash of its children. Merkle trees enable efficient and secure verification of large datasets by allowing parties to confirm the integrity of specific data blocks without needing to examine the entire dataset. This property makes Merkle trees invaluable for securing data integrity in distributed systems, blockchains, and peer-to-peer networks. In the context of PQC, hash-based cryptography provides several advantages. Firstly, hash functions are relatively simple and efficient, making them suitable for resource-constrained environments. Additionally, hash-based cryptographic schemes often rely on mathematical problems that are not believed to be vulnerable to quantum attacks, offering a robust foundation for secure communication and data integrity in a post-quantum world.

Despite its strengths, hash-based cryptography also has limitations and challenges. One potential concern is the scalability of hash-based schemes for certain applications, particularly in scenarios where a high volume of signatures or verifications is required. Additionally, as with any cryptographic system, the security of hash-based schemes depends on the underlying hash function. Therefore, ongoing research and evaluation of hash function designs and implementations are essential to maintain their security in the face of evolving threats. The hash-based cryptography represents a promising avenue for addressing the challenges posed by quantum computing to traditional cryptographic systems. By leveraging the security properties of hash functions, hash-based cryptographic schemes offer robust solutions for digital signatures, data integrity verification, and secure communication in the era of post-quantum cryptography. As research in this field continues to advance, hash-based cryptography is poised to play a significant role in shaping the future of secure digital communication and information exchange.

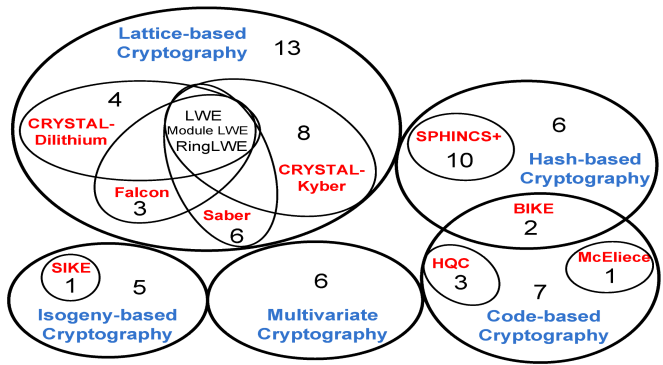

Fig 3. Types of PQC

VII. CASE STUDIES:

In recent years, the emergence of quantum computing has prompted intensive research into Post-Quantum Cryptography (PQC) as a potential solution to secure sensitive data and communications in the face of quantum threats. Several case studies and real-world applications have emerged, shedding light on the practical implementation and efficacy of PQC in various domains.

*A) Government and Defense:* A government agency responsible for national defense strategically integrates Post-Quantum Cryptography (PQC) algorithms into its secure communication systems to fortify its capabilities against emerging quantum threats. By adopting lattice-based or code-based PQC schemes, the agency aims to encrypt classified information and establish secure communication channels resistant to quantum attacks. This proactive approach underscores the agency's commitment to safeguarding sensitive data and ensuring the integrity of communications amidst evolving technological landscapes. However, the transition to PQC presents significant challenges, including the need for substantial investments in infrastructure upgrades and specialized training to seamlessly integrate PQC algorithms with existing systems. Despite these hurdles, the agency recognizes the imperative of embracing PQC as a cornerstone of its cybersecurity strategy, prioritizing resilience against quantum-enabled adversaries.

While specific case studies within government agencies may not be readily available due to the classified nature of defense operations, hypothetical examples and ongoing research initiatives underscore the growing importance of PQC in national security frameworks. Governments worldwide are increasingly acknowledging the urgency of quantum-safe cryptography and are engaging in collaborative efforts with researchers and industry stakeholders to drive innovation and ensure effective PQC deployment. As quantum computing capabilities evolve, proactive measures to adopt PQC protocols become paramount, signaling a paradigm shift in cryptographic standards and strategies aimed at mitigating quantum-enabled risks across critical infrastructure sectors.

*B) Financial Sector:* A leading financial institution recognizes the escalating risks posed by quantum computing and strategically adopts Post-Quantum Cryptography (PQC) solutions to fortify the security of its online banking and payment platforms. Implementing hash-based or multivariate-based PQC algorithms, the institution aims to safeguard customer transactions and sensitive financial data from potential quantum attacks. By embracing PQC measures, the financial entity demonstrates its commitment to maintaining the integrity of its digital infrastructure and protecting customers against evolving cyber threats. This proactive approach not only enhances the institution's cybersecurity posture but also strengthens customer trust and confidence in its services, emphasizing a dedication to robust data protection in an increasingly interconnected financial ecosystem.

However, amidst the pursuit of enhanced security, financial organizations encounter formidable challenges in implementing PQC solutions effectively. Ensuring the scalability and performance of PQC systems while adhering to stringent regulatory compliance standards presents significant hurdles for institutions operating within highly regulated environments. Moreover, the complex interplay between PQC algorithms and existing cryptographic protocols necessitates careful evaluation and testing to mitigate potential compatibility issues and ensure seamless integration. Despite these challenges, financial institutions remain steadfast in their commitment to adopting PQC as a cornerstone of their cybersecurity strategy, recognizing the imperative of staying ahead of emerging threats and safeguarding the integrity of financial transactions in an era of quantum-enabled risks.

*C) Healthcare Industry:* In response to the increasing vulnerability of electronic health records (EHRs) and patient information to cyber threats, a healthcare provider strategically implements Post-Quantum Cryptography (PQC) protocols to fortify its data security measures. By integrating quantum-resistant signature schemes or symmetric encryption algorithms into its systems, the healthcare provider aims to establish robust defenses against potential breaches and unauthorized access to sensitive medical data. Through the adoption of PQC technologies, the organization demonstrates its commitment to safeguarding patient privacy and confidentiality, prioritizing the integrity and security of healthcare information in an era of evolving cyber risks and quantum-enabled threats.

However, the deployment of PQC in healthcare settings presents unique challenges that must be addressed to ensure effective implementation. Balancing the computational overhead of PQC algorithms with the resource constraints of healthcare IT infrastructure requires meticulous optimization and testing processes. Healthcare providers must navigate the complexities of integrating PQC solutions without compromising system performance or impeding critical workflows. Moreover, ensuring interoperability and compatibility with existing systems and protocols adds layers of complexity to the deployment process. Despite these challenges, healthcare organizations remain dedicated to leveraging PQC technologies as integral components of their cybersecurity strategies, recognizing the imperative of safeguarding patient data and maintaining trust in the confidentiality and integrity of healthcare information.

Through ongoing research, collaboration, and innovation, healthcare providers strive to overcome implementation barriers and embrace PQC as a cornerstone of secure and resilient healthcare systems.

## VIII. Analysis of Effectiveness of PQC and challenges faced

Post-Quantum Cryptography (PQC) is a critical area of study and development in the realm of cybersecurity, particularly in safeguarding sensitive data and communications in the face of potential threats posed by quantum computers. Here's an analysis of the effectiveness of PQC in protecting sensitive data and communications, along with challenges faced and lessons learned from its deployment:

*A) Effectiveness of PQC:*

1. Resistance Against Quantum Attacks: Traditional cryptographic algorithms like RSA and ECC are vulnerable to attacks from quantum computers due to their reliance on integer factorization and discrete logarithm problems, which can be solved efficiently by quantum algorithms like Shor's algorithm. PQC algorithms, on the other hand, are designed to resist attacks even in the presence of quantum computing capabilities.

2. Security Guarantees: PQC algorithms provide security guarantees based on mathematical problems that are believed to be hard even for quantum computers. These algorithms rely on problems such as lattice-based cryptography, code-based cryptography, hash-based cryptography, and multivariate polynomial cryptography, which are not easily solved by quantum algorithms.

3. Compatibility and Integration: PQC algorithms are being standardized and integrated into existing cryptographic protocols and systems to ensure seamless adoption and compatibility with current infrastructure. Efforts by organizations like NIST (National Institute of Standards and Technology) are instrumental in developing standards for PQC algorithms.

*B) Challenges Faced:*

1. Performance Overhead: Many PQC algorithms tend to have higher computational and communication overhead compared to traditional cryptographic algorithms. This can pose challenges, especially in resource-constrained environments and applications where efficiency is crucial.

2. Interoperability and Transition: Integrating PQC algorithms into existing systems while maintaining interoperability and backward compatibility is complex. It requires careful planning and may involve transitional periods during which both traditional and PQC algorithms need to be supported.

Algorithm Maturity and Standardization: While some PQC algorithms have reached a certain level of maturity, others are still in the research or development phase. Standardization efforts are ongoing, and the selection of suitable algorithms requires thorough evaluation and consideration of various factors such as security, efficiency, and scalability.

## IX. Future Scope

Future directions in post-quantum cryptography (PQC) research are poised to drive significant advancements in cryptography theory and practice. One promising direction involves the exploration of quantum-resistant cryptographic primitives and protocols that can withstand the computational capabilities of quantum adversaries. Researchers are actively investigating novel mathematical constructs and algorithmic techniques to develop robust cryptographic schemes that offer security guarantees in the presence of quantum attacks. Additionally, the standardization and validation of PQC algorithms represent crucial avenues for advancing the adoption and deployment of quantum-resistant cryptography in real-world applications. Collaborative efforts among academia, industry, and standardization bodies are instrumental in establishing rigorous evaluation criteria and certification frameworks to ensure the reliability and interoperability of PQC solutions across diverse computing environments.

Furthermore, the integration of PQC algorithms into emerging technologies and applications presents new opportunities and challenges for cybersecurity and digital privacy. As the Internet of Things (IoT), cloud computing, and distributed ledger technologies continue to proliferate.

The need for scalable and efficient PQC solutions becomes increasingly pronounced. Future research endeavors aim to address scalability concerns and optimize the performance of PQC algorithms in resource-constrained environments, enabling seamless integration into next-generation cryptographic architectures. Moreover, interdisciplinary collaborations between cryptography, quantum computing, and information security domains hold the promise of unlocking innovative solutions and shaping the trajectory of post-quantum cryptography research. By embracing these future directions, the cybersecurity community can harness the transformative potential of PQC to fortify the resilience and integrity of digital communication systems in the quantum era.

## X. Conclusion

In conclusion, the research underscores the pivotal role of post-quantum cryptography (PQC) in fortifying digital communication against the looming threat of quantum-enabled attacks. With the advent of quantum computing on the horizon, the imperative to transition towards quantum-resistant cryptographic algorithms has become paramount. Through a comprehensive examination of the vulnerabilities inherent in classical cryptographic systems, the study highlights the pressing need for robust PQC solutions capable of withstanding the computational power of quantum adversaries. Furthermore, by delineating the anticipated advancements and potential impact of PQC on the future of cybersecurity, the research underscores the significance of embracing quantum-resistant cryptographic primitives and protocols. As we navigate the complexities of the quantum era, the adoption and standardization of PQC algorithms emerge as indispensable components in safeguarding the confidentiality, integrity, and authenticity of digital communication in an increasingly interconnected world. In essence, the research reaffirms the critical importance of PQC in shaping the future landscape of cybersecurity and underscores its transformative potential in mitigating the risks posed by quantum computing advancements.